\documentclass[a4paper,12pt]{revtex4}                    
\usepackage{amssymb,amsthm,latexsym}
\newcommand{\Med}[1]{\left\langle #1 \right\rangle}
\newcommand{\med}[1]{\langle #1 \rangle}
\newcommand{\toro}{T_{L}}
\newcommand{\kac }{\lim_{\gamma\to0}\lim_{L\to\infty}}
\newtheorem{theorem}{Theorem}

\newtheorem{lemma}{Lemma}
\begin{document}

\title{Finite-range spin glasses in the Kac limit: free energy and local observables}

\author{Silvio Franz~\cite{sf},
Fabio Lucio Toninelli~\cite{flt}}

\address{ \cite{sf} The Abdus Salam
International Center for Theoretical Physics, Condensed Matter Group\\
Strada Costiera 11, P.O. Box 586, I-34100 Trieste, Italy\\
\cite{flt} Institut f\"ur Mathematik, Universit\"at Z\"urich\\
Winterthurerstrasse 190, CH-8057 Z\"urich, Switzerland
}

\date{\today}

\begin{abstract}
 We study a finite range spin glass model in arbitrary dimension,
where the intensity of the coupling between spins decays to zero over
some distance $\gamma^{-1}$.  We prove that, under a positivity condition for 
the interaction potential, the infinite-volume free energy of the
system converges to that of the Sherrington-Kirkpatrick model, in the
Kac limit $\gamma\to0$. We study the implication of this convergence
for the local order parameter, {\em i.e.}, the local overlap distribution function
and a family of susceptibilities to it associated, and we show that
locally the system behaves like its mean field analogue. Similar results are obtained for models with
$p$-spin interactions.
Finally, we discuss a possible approach to the problem of the existence of long range order
for finite $\gamma$, based on a large deviation functional for overlap profiles. 
This will be developed in future work.
\end{abstract}

\maketitle

\section{Introduction}

Since their first introduction by van der Waals in 
the second half of the 19th century, mean field theories offer a simplified setting to
understand the complex collective phenomena underlying phase
transitions and low temperature ordering. These theories, it appeared
very soon, are plagued by several pathologies, which can be traced back to 
the fact that  the finite range character of the interactions is neglected.  For
example,  mean field theory predicts the possibility of low
temperature ordering, independently of the space dimensionality. In
addition the free energy, in principle convex in extended system, can
present non-convexities at mean field level. 
In a series of classical papers Kac,  Uhlenbeck and Hemmer \cite{kuh} stressed the
role of the interaction range in these pathologies. Using
a one-dimensional model for liquid-vapor transition they showed that
in the so-called Kac limit \cite{kac}, when range of interaction $\gamma^{-1}$ is sent to
infinity {\it after} the thermodynamic limit, while the total interaction strength is kept constant,
one recovers the van der
Waals theory complemented by the Maxwell construction, thus
eliminating the unphysical non-convexity of the thermodynamic
potential.  It was also shown that coherently with general principles,
while the free energy for finite $\gamma$ is close to the
corresponding mean field value, the phase transition only appears
for $\gamma^{-1}\to\infty$.  The comprehension of the question was
considerably extended by Lebowitz and Penrose \cite{lp}, who could study the Kac limit 
in great generality for any
value of the spatial dimension, confirming the validity of mean field theory with Maxwell
construction. The last decade has seen a renewed
interest in Kac models, which have been used as a starting point for
rigorous expansions around mean field. Remarkable progress have been
achieved in the analysis of models with large but finite range of interaction, both for systems
without quenched disorder (ferromagnets \cite{cp} \cite{bz} and liquid models \cite{lpres1}) and with
quenched disorder as the Random Field Ising Model \cite{RFIM} and the Hopfield model \cite{hopfield}.

A case where the application of the mean field theory to spatially
extended systems is particularly controversial is that of spin
glasses.  In that case, the mean field theory based on the Parisi
solution \cite{MPV} of the long-range interaction model of Sherrington and
Kirkpatrick (SK) \cite{sk}, predicts a low temperature glassy phase with ergodicity breaking, not
 associated to any physical symmetry breaking. In particular, there is a 
transition to a low temperature non-ergodic phase even in
presence of a field breaking explicitly the spin reversal symmetry of
the Hamiltonian. This results in an organization of the low free energy
states which gives rise to complex statistical properties of the so-called
overlap distribution function, describing the probability law, induced by the Gibbs measure and
the quenched disordered couplings, of the scalar product among configurations. 
This picture, correct for the mean field SK model, has been challenged, in the case of finite dimensional 
models, by
phenomenological theories based on scaling arguments -the droplet
picture \cite{drop1} \cite{drop2} of the spin-glass phase- that predict the absence of Parisi
ordering in any finite dimension. According to this point of view,  the spin-glass
transition just corresponds to spin-reversal symmetry breaking
and consequently cannot be present in a magnetic
field. 
Years of theoretical
debate, experimental results, accurate numerical simulations and even
rigorous arguments \cite{ns} have failed to give a conclusive answer to the
question.

It is natural in this context to look at spin glasses with Kac-type 
interactions, as a tool to study the relation between mean field and
finite range. Spin glasses with Kac interactions were to our knowledge
defined in \cite{froelich}, where it was shown that their
free energy converges in the Kac limit to the mean field value for sufficiently high temperature, and later considered
in \cite{bovier-kac}, where the result was extended up to the critical temperature $T_c=1$ of the 
SK model. Little progress was made until the introduction of
interpolating techniques by Guerra and Toninelli in the rigorous study
of disordered systems. Through those methods, it was first proven that
the Kac free energy is, for a large class of Kac potentials, bounded
from below by the one of the SK model \cite{gt}. Finally, joining
the interpolating technique with the idea of dividing the system into large blocks
where it essentially behaves like the mean field one \cite{lp}, 
it was possible to show the convergence of the free energy to the SK
one for all temperature \cite{ft}.

The scope of this paper is to give full details of the proofs of the
results in \cite{gt,ft} and to discuss the implication of the Kac
limit for local quantities. In particular we will discuss the
properties of the local overlap probability distribution and local
susceptibilities and show that for small $\gamma$ they are close
respectively to the overlap probability distribution and susceptibilities of the SK model at the
same temperature.
As a further extension, we generalize these results to the case of Kac spin glass models with $p$-body interactions,
where $p$ is an even integer.

The present work is organized as follows: in Section \ref{sec:def} we define the models and state the main results.
In Section \ref{sec:ft} we prove convergence of the free energy to the mean field limit, when $\gamma\to0$,
while in Section \ref{sec:lo} we prove convergence for the distribution of the local overlaps.
Finally, in Section \ref{sec:out} we outline a possible strategy to study the occurrence of long-range order
in finite range spin glasses, when $\gamma$ is small but finite. This requires the introduction of a large 
deviation functional for the profile of two replicas. 
This strategy will be pursued in a forthcoming paper \cite{FDG}.

\section{The models and the main results}
\label{sec:def}

\subsection{Spin glasses with Kac-type interactions}

In this section we define a finite range version of the $p$-spin spin
glass model well suited to study the Kac limit. We remind that the 
original infinite range model is defined \cite{derrida}, 
for an integer $p$ and some magnetic field $h\in{\mathbb R}$,  
by a set of $N$ 
Ising spins $\sigma_i=\pm1$, $i=1,...,N$ interacting via the Hamiltonian: 
\begin{equation}
\label{hP}
H_N^{(p)}(\sigma,h;J)=-\sqrt{\frac{p!}{2N^{p-1}}}\sum_{1\le i_1<\cdots< i_{p}\le N}
J_{i_1\cdots i_{p}}
\sigma_{i_1}\cdots\sigma_{i_{p}}
-h\sum_{i=1}^N\sigma_i,
\end{equation}
where the couplings $J_{i_1\cdots i_p}$ are 
independent identically distributed (i.i.d.) Gaussian random variables
with averages
\begin{equation}
E J_{i_1\cdots i_p}=0\hspace{1cm} E J^2_{i_1\cdots i_p}=1. 
\label{av}
\end{equation}
For $p=2$ one recovers the usual SK model.

We propose a generalization of the model to the $d$-dimensional
lattice $\mathbb{Z}^d$.  Throughout all the paper we will assume
periodic boundary conditions for convenience, in order to ensure translation invariance.  Therefore, we will
always consider the system on $\toro$, the $d$-dimensional discrete
torus of side $L$ and cardinality $N=L^d$.  Given $\gamma>0$ and a family
$J_{i_1\cdots i_p}$, $i_r\in\toro, r=1,\cdots,p$ of i.i.d. Gaussian random variables with averages as in
(\ref{av}), we define the finite volume Hamiltonian for the $p$-spin
spin glass with Kac-type interactions as
\begin{eqnarray}
\label{Hp}
H^{(p,\gamma)}_L(\sigma,h;J)&=&  -K^{(p,\gamma)}_L(\sigma;J)-h\sum_{i\in\toro}\sigma_i\\
\nonumber
&=&
-\sum_{i_1,\cdots,i_p\in \toro}\sqrt {w^{(p)}(i_1,\cdots,i_p;
\gamma)}
J_{i_1\cdots i_p}\sigma_{i_1}\cdots\sigma_{i_p}-h\sum_{i\in\toro}\sigma_i.
\end{eqnarray}
Here, 
\begin{eqnarray}
\label{wij}
w^{(p)}(i_1,\cdots,i_p;\gamma)
=
\frac{\sum_{k\in \toro}\psi(\gamma|i_1-k|)\cdots
\psi(\gamma|i_p-k|)}{W(\gamma)^{p/2}}
\end{eqnarray}
and
\begin{equation}
W(\gamma)=\left(\sum_{k\in \toro}\psi(\gamma|k|)\right)^2,
\end{equation}
where $\psi(|x|)$, $x\in {\mathbb R}^d$, is a  non-negative
function with compact support,
$$
\psi(|x|)=0\;\;\; \mbox{if}\;\;\; |x|\ge1,
$$
sufficiently regular to be Riemann integrable.

For $\gamma\simeq 0$, it is immediate to see that one recovers the usual form \cite{lp} 
for the Kac potential, {\em i.e.},
\begin{equation}
w^{(2)}(i,j;\gamma)\simeq \gamma^d\phi(\gamma|i-j|),
\end{equation}
where in our case 
\begin{equation}
\phi(|i-j|)=\frac{\int \psi(|i-k|)\psi(|j-k|)d^dk }{\left(\int \psi(|k|)d^dk \right)^2}.
\end{equation}
On the other hand, the reason for the particular choice (\ref{wij}) is 
that it guarantees that $w^{(2)}$ is positive definite, {\em i.e.},
\begin{equation}
\sum_{i,j\in\toro} w^{(2)}(i,j;\gamma)v_i v_j\ge0\;\;\;\forall\;
\{v_i\}_{i\in\toro}, v_i\in{\mathbb R},
\end{equation}
and that a suitable positive definiteness property, implied by Lemma \ref{sobolev} below, is satisfied by
$w^{(p)}$, for any even $p$. This property will turn out to be essential in proving our results.

It is easy to realize that the potentials $w^{(p)}$ 
satisfy the following properties:
\begin{enumerate}
\item invariance with respect to translations on the torus:
\begin{equation}
w^{(p)}(i_1,\cdots,i_p;\gamma)=w^{(p)}(i_1+k,\cdots,i_p+k;\gamma)\;\;\forall 
k\in\toro
\end{equation}
\item finite range of order $\xi=1/\gamma$:
\begin{equation}
w^{(p)}(i_1,\cdots,i_p;\gamma)=0\;\; \mbox{if}\;\;\exists\; a,b:
|i_a-i_b|\ge 2\xi
\end{equation}
\item normalization:
\begin{equation}
\sum_{i_2,\cdots,i_p\in\toro} w^{(p)}(i_1,\cdots,i_p;\gamma)=
1\;\;\forall i_1\in\toro,
\end{equation}
and
\begin{equation}
\label{consistenza}
\sum_{i_{r+1},\cdots,i_p\in\toro} w^{(p)}(i_1,\cdots,i_p;\gamma)=
w^{(r)}(i_1,\cdots,i_r;\gamma)
\;\;\forall i_1,\cdots,i_{r}\in\toro,
\end{equation}
for $1<r<p$, 
\end{enumerate}
besides of course symmetry with respect to index permutation. 
Note that $w^{(1)}(i;\gamma)=1$. Note also 
that, while $w^{(p)}(i_1,\cdots,i_p;\gamma)$ 
in principle depends on the size of the
system, this dependence disappears as soon as $L>1/\gamma$ and is therefore
inessential in view of the fact that we consider the limit  $\gamma\to0$
only after $L\to\infty$.

{\bf Remark} 
All the results of the present paper extend to the case where $\psi(|x|)$ is 
only assumed to decay sufficiently fast for $|x|\to\infty$ so that
\begin{equation}
\psi(|x|)\le C |x|^{-d-\delta},
\end{equation}
for some $\delta,C>0$. A similar condition was required in \cite{lp}.

For a given inverse temperature $\beta$, we denote 
as $Z^{(p,\gamma)}_L(\beta,h;J)$ the disorder 
dependent partition function of the model (\ref{Hp}), by $\med.$ the corresponding Gibbs average, 
and by $f_L^{(p,\gamma)}(\beta,h)$ the finite volume quenched free energy
\begin{equation}
f_L^{(p,\gamma)}(\beta,h)=-\frac1{\beta L^d}E \ln Z^{(p,\gamma)}_L(\beta,h;J).
\end{equation}
Our first result shows that, as a generalization of \cite{gt} and \cite{ft}, the free energy of the 
Kac model tends, for $\gamma\to0$, to that of the corresponding
mean field model. Let us first of all
redefine the Hamiltonian (\ref{hP}) in a slightly different way, in order to make the comparison 
with model (\ref{Hp}) more straightforward:
\begin{equation}
\label{hMF}
H_L^{(p)}(\sigma,h;J)= 
-K^{(p)}_L(\sigma;J)-h\sum_{i\in\toro}\sigma_i=-\sum_{i_1,\cdots, i_{p}\in\toro}
\frac{J_{i_1\cdots i_{p}}}
{L^{d(p-1)/2}}
\sigma_{i_1}\cdots\sigma_{i_{p}}
-h\sum_{i\in \toro}\sigma_i,
\end{equation}
whose partition function will be denoted as $Z^{(p)}_L(\beta,h;J)$.
Of course, in this case the geometrical structure of the lattice and the space dimensionality are completely
inessential. Then, the following holds:
\begin{theorem}
\label{generalizz}
For any $\beta,h$ and $p$ even, and for any choice of the
function $\psi$ in (\ref{wij}), the following limit exists and satisfies
\begin{eqnarray}
\lim_{\gamma\to0}f^{(p,\gamma)}(\beta,h)\equiv
\lim_{\gamma\to0}\lim_{L\to\infty} f_L^{(p,\gamma)}(\beta,h)=f^{(p)}(\beta,h),
\end{eqnarray}
where $f^{(p)}(\beta,h)$ is the infinite-volume quenched free
energy for the mean field $p$-spin spin glass with Hamiltonian (\ref{hMF}), i.e., 
\begin{equation}
\label{limtermMF}
f^{(p)}(\beta,h)=\lim_{L\to\infty}f_L^{(p)}(\beta,h)\equiv 
-\lim_{L\to\infty}\frac1{\beta L^d}E\ln Z^{(p)}_L(\beta,h;J).
\end{equation}
\end{theorem}

{\bf Remark } The existence of the infinite-volume free energy
$f^{(p,\gamma)}(\beta,h)$ for the model (\ref{Hp}) can be proven along
the lines of \cite{vanenter}, while for the mean field model
(\ref{hMF}) the analogous result was proven in \cite{limterm}. It is
interesting to remark that the methods we develop in the present paper
can be employed to obtain a new proof of the existence of the limit
(\ref{limtermMF}), see end of Section \ref{sec:ft}.

\subsection{Local observables}

The next natural question after proving the convergence of the free
energy to its mean field limit, is to investigate in which sense the
system has a behavior close to mean field for small but finite
$\gamma$. In particular we would like to have informations on
the overlap distribution function and its associated susceptibilities, 
whose non-trivial behavior characterize the Parisi order in mean
field theory.  As a first step we investigate the behavior of 
``local'' observables on the scale $1/\gamma$, for $\gamma$ small. We
will show that on this scale the behavior of the system is close to
that of the mean field model, in any dimension.  The natural approach is to add
suitable perturbations to the Hamiltonian, such that the averages of
the observables of interest are given by the derivatives of the free
energy with respect to the perturbing parameters.

Given two configurations $\sigma^1,\sigma^2$ of the system and $k\in\toro$, let us define the ``local overlap''
\begin{equation}
\label{ovloc}
q_k^\gamma(\sigma^1,\sigma^2)=
\sum_{i\in\toro}\frac{\psi(\gamma|i-k|)}{W(\gamma)^{1/2}}
\sigma^1_i\sigma^2_i.
\end{equation}
In particular we will have in mind the case where $\psi$ is the Heaviside function 
\begin{equation}
\label{speciale}
\psi(|x|)=1_{\{|x|_\infty\le1\}},
\end{equation}
$|x|_\infty$ being defined as $\max_{r=1,\cdots,d}|x_r|$. In this case,
$W(\gamma)\simeq(2/\gamma)^{2d}$ and Eq. (\ref{ovloc}) reduces to 
the familiar definition
\begin{equation}
q_k^\gamma(\sigma^1,\sigma^2)=
\left(\frac\gamma2\right)^{d}\sum_{i\in\toro:|i-k|_\infty\le 1/\gamma}
\sigma^1_i\sigma^2_i
\end{equation}
of the overlap in the cube of side $2\gamma^{-1}$ centered at the site $k$.
The Gibbs measure and the quenched couplings induce the probability 
distribution
 $P^{(p,\gamma)}_{L}(q)$ for the overlap, depending also on $\beta$ and $h$, which we write 
formally as
\begin{eqnarray}
P^{(p,\gamma)}_{L}(q) =E\left(\frac{\sum_{\sigma^1,\sigma^2} \exp\left(-\beta
  H^{(p,\gamma)}_L(\sigma^1,h;J)-\beta H^{(p,\gamma)}_L(\sigma^2,h;J)\right)
 \delta(q_k^\gamma(\sigma^1,\sigma^2)-q)}{\left(Z^{(p,\gamma)}_L(\beta,h;J)\right)^2}\right). 
\label{ploc}
\end{eqnarray} 
As in \cite{ghigu} we consider a family of perturbations associated to the
function (\ref{ploc}), and introduce the Hamiltonian
\begin{equation}
\label{hkac}
H^{(p,\gamma)}_L(\sigma,\{\varepsilon\},h;J)=
-K^{(p,\gamma)}_L(\sigma;J)-\sum_{r\ge1}\varepsilon_{r}
K_L^{(r,\gamma)}(\sigma;J^{(r)})-h\sum_{i\in \toro}\sigma_i,
\end{equation}
where the families of Gaussian random variables $J^{(r)}$ are independent for different $r$, and 
the real numbers $\varepsilon_r$ decay to zero sufficiently
rapidly when $r\to\infty$, to ensure that the corresponding free energy 
is finite.
Actually, we will require that $\{\varepsilon_r\}_{r\ge1}$ belongs to a region $R_p\subset {\mathbb R}^\infty$,
defined precisely in Appendix \ref{app_epsilon},  characterized by the fact that
not only $|\varepsilon_r|$ are sufficiently small for $r$ large, but also that
\label{talagrandata}
\begin{equation}
|\varepsilon_s|\ll |\varepsilon_2|, \;s\ge3.
\end{equation}
In any case, we will be eventually interested in letting {\em all} the $\varepsilon_r$'s tend to zero,
so that the corresponding terms in the Hamiltonian are actually small
perturbations.  

Let $P_{L,\varepsilon}^{(p,\gamma)}(q)$ be the
probability 
distribution density of $q_{12}^\gamma\equiv q^\gamma_0(\sigma^1,\sigma^2)$ in presence of the
perturbations.  The role of the perturbations in (\ref{hkac}) is
easily understood. Indeed, we will see that
\begin{equation}
  \partial_{\varepsilon_r}f^{(p,\gamma)}_L(\beta,\{\varepsilon\},h)=
-\beta\varepsilon_r(1-E\med{(q^\gamma_{12})^r})=
-\beta\varepsilon_r\left(1-\int dq\,P_{L,\varepsilon}^{(p,\gamma)}(q)\,q^r\right),
\end{equation}
where $f^{(p,\gamma)}_L(\beta,\{\varepsilon\},h)$ is the finite volume free energy of the model (\ref{hkac}).
 Then, one has
\begin{theorem}
\label{teooverlap}
For any choice of $\psi$ in (\ref{wij}), and for almost every value of
$\{\varepsilon\}$,
\begin{equation}
\label{convergenza}
\kac P^{(p,\gamma)}_{L,\varepsilon}(q)=P^{(p)}_{\varepsilon}(q),
\end{equation}
where the r.h.s. is the infinite volume quenched average of the
probability distribution of the full overlap
\begin{equation}
\label{overlap}
q_{12}=L^{-d}\sum_{i\in\toro}\sigma^1_i\sigma^2_i
\end{equation}
for the ``perturbed p-spin mean field model'' defined
by the Hamiltonian
\begin{equation}
\label{hSK}
H^{(p)}_L(\sigma,\{\varepsilon\},h;J)=-K^{(p)}_L(\sigma;J)-\sum_{r\ge
1}\varepsilon_{r}
K_L^{(r)}(\sigma;J^{(r)})-h\sum_{i\in \toro}\sigma_i.
\end{equation}
\end{theorem}
{\bf Remarks} Theorem \ref{teooverlap} shows that for small but finite 
$\gamma$,
on the scale of the range of the interaction, the overlap 
probability distribution is close to the corresponding 
mean field one at the same
temperature, which is known to be  non-trivial at low temperature. 
Of course, as it already happens in non-disordered models, this
does not have implications on the possible presence of long range order.

As it is well known \cite{MPV}, at the mean field level the
probability distribution of a single overlap is not sufficient to give
a full description of a spin glass model, and the knowledge of the
joint distribution $P(\{q_{ab}\})$ of the overlaps $q_{ab}$ among any
two replicas $a,b$ is required.  Parisi's theory for the mean field
model predicts a highly non-trivial {\em ultrametric} structure for
$P(\{q_{ab}\})$, whose validity has not been rigorously established
yet. It would be possible to extend the ideas of the present section, and to show that the 
joint distribution of the local overlaps between several replicas behaves, for $\gamma\to0$, 
like the corresponding one for the mean field model, but we will 
not pursue this way. In this case, the 
perturbations to be added are the Kac-type analogs of those introduced in \cite{fmpp1}. 

Finally, let us remark that the convergence in (\ref{convergenza}) is proved to hold only 
for {\em almost every} $\{\varepsilon\}$, so that nothing can be said {\em a priori} 
for $\varepsilon\equiv0$, corresponding to the original model (\ref{Hp}), since the Gibbs averages 
need not be continuous with respect to the perturbing parameters. The same problem arises also at the level
of mean field theory (see for instance \cite{talaAC}), 
and is a drawback of the the method of stochastic perturbations.

\section{Kac limit for the free energy}

\label{sec:ft}

{\em Proof of Theorem \ref{generalizz}} 

First of all, extending the ideas of \cite{gt}, we will prove the lower bound
\begin{equation}
\label{lb}
f^{(p,\gamma)}_L(\beta,h)\ge f^{(p)}_L(\beta,h),
\end{equation}
uniformly in $\gamma>0$ and $L>1/\gamma$. To this purpose, let 
\begin{eqnarray}
\label{z1}
Z_L(t)=\exp\beta\left(\sqrt t K^{(p,\gamma)}_L(\sigma;J)+
\sqrt{1-t}K^{(p)}_L(\sigma;J')+
h\sum_{i\in\toro}\sigma_i\right),
\end{eqnarray}
where the families of Gaussian variables $J$ and $J'$ are mutually independent.
Then, 
\begin{equation}
-\frac1{\beta L^d}E \ln Z_L(0)=f^{(p)}_L(\beta,h)
\end{equation}
and
\begin{equation}
-\frac1{\beta L^d}E \ln Z_L(1)=f^{(p,\gamma)}_L(\beta,h)
\end{equation}

The $t$-derivative of the free energy 
can be written, introducing two replicas $\sigma^1,\sigma^2$ of the system with the same 
disorder realization, as
\begin{eqnarray}
\label{deriv_lb}
-\frac d{dt}\frac1{\beta L^d}E \ln Z_L(t)
=\frac\beta{2}
E\left(\sum_{i_1,\cdots,i_{p}\in\toro}
\frac{w^{(p)}(i_1,\cdots,i_{p};\gamma)}{L^d}
\med{\sigma^1_{i_1}\sigma^2_{i_1}\cdots\sigma^1_{i_p}\sigma^2_{i_{p}}}-\med{q_{12}^{p}}\right),
\end{eqnarray}
where $q_{12}$ is the overlap defined in (\ref{overlap}) and the $t$-dependence is implicit in the Gibbs average
$\med.$.
Thanks to Lemma \ref{sobolev} below and to the translation invariance
of the system, ensured by the  periodic boundary conditions, (\ref{deriv_lb}) equals
\begin{eqnarray}
\label{derivata}
&&\frac\beta2
\sum_{r=2}^{p} \left(\begin{array}l
p\\
r\\
\end{array}
\right)
E\Med{q_{12}^{p-r}
\left(\sum_{i\in\toro}\frac{\psi(\gamma|i|)}
{W(\gamma)^{1/2}}
(\sigma^1_{i}\sigma^2_{i}-q_{12})\right)^r
}.
\end{eqnarray}
Now, it is immediate to see that
\begin{equation}
\sum_{r=2}^{p} \left(\begin{array}l
p\\
r\\
\end{array}
\right)x^{p-r}y^r=(x+y)^{p}-x^{p}-p x^{p-1}y\ge0
\end{equation}
for any $x,y\in {\mathbb R}$ and $p$ even, so that the derivative in (\ref{derivata}) 
is non-negative, and (\ref{lb}) follows.

On the other hand, the idea to obtain the upper bound
\begin{equation}
\label{ub}
\limsup_{\gamma\to0} f^{(p,\gamma)}(\beta,h)\le f^{(p)}(\beta,h)
\end{equation}
is to interpolate between the Kac model in the box  $\toro$ and
a system made of a collection of many independent mean field subsystems enclosed in cubes of
side $\ell$ \cite{ft} . The crucial point, as in \cite{lp}, is to choose
\begin{equation}
\label{scale}
1\ll \ell \ll 1/\gamma \ll L,
\end{equation}
and to let  the  three lengths diverge in this order.
Let us divide  $\toro$ into
sub-cubes $\Omega_n$ of side  $\ell$, $n=1,\cdots,(L/\ell)^d$, 
and introduce the partition function
\begin{eqnarray}
\label{z2}
Z_L(t)&=&\sum_{\sigma}\exp\beta\left(\sqrt t
K_L^{(p,\gamma)}(\sigma;J)+\sqrt{1-t}\sum_n \sum_{i_1,\cdots,i_p\in \Omega_n}\frac{J'_{i_1\cdots i_p}}
{\ell^{d(p-1)/2}}
+h\sum_{i\in\toro}\sigma_i\right),
\end{eqnarray}
which interpolates between the Kac model and a collection of many non-interacting mean field models in the different 
boxes, with independent couplings.
In this case, 
\begin{equation}
-\frac1{\beta L^d}E \ln Z_L(0)=f^{(p)}_\ell(\beta,h)
\end{equation}
and
\begin{equation}
-\frac1{\beta L^d}E \ln Z_L(1)=f^{(p,\gamma)}_L(\beta,h)
\end{equation}
while the $t$-derivative of the free energy is
\begin{eqnarray}
\label{deriv_ub}
\nonumber
-\frac d{dt}\frac1{\beta L^d}E \ln Z_L(t)
=\frac\beta{2}E
\left(\sum_{i_1,\cdots,i_{p}\in\toro}
\frac{w^{(p)}(i_1,\cdots,i_{p};\gamma)}{L^d}
\med{\sigma^1_{i_1}\sigma^2_{i_1}\cdots\sigma^1_{i_p}\sigma^2_{i_{p}}}
-\left(\frac\ell L\right)^d \sum_n \med{(q^{(n)}_{12})^{p}}\right),
\end{eqnarray}
where 
$$
q^{(n)}_{12}=\frac1{\ell^d}\sum_{i\in \Omega_n}\sigma^1_i\sigma^2_i
$$
is the partial overlap referring to the $n$-th box.
Defining 
\begin{equation}
w^{(p)}_+(n_1,\cdots,n_p;\gamma)=\max_{i_r\in\Omega_{n_r},r=1,\cdots,p}w^{(p)}(i_1,\cdots,i_p;\gamma),
\end{equation}
one has the immediate bound 
\begin{eqnarray}
\label{bound_deriv_ub}
-\frac d{dt}\frac1{\beta L^d}E \ln Z_L(t)
&\le&\frac\beta{2}\left(\frac\ell L\right)^d E
\left(\ell^{d(p-1)}\sum_{n_1,\cdots,n_{p}}
w_+^{(p)}(n_1,\cdots,n_{p};\gamma)
\med{q_{12}^{(n_1)}\cdots q_{12}^{(n_p)}}\right.\\\nonumber
&&\left.-\sum_n \med{(q^{(n)}_{12})^{p}}\right)
\end{eqnarray}
and, thanks to Lemma \ref{lemmaLP} below,
\begin{eqnarray}
-\frac d{dt}\frac1{\beta L^d}E \ln Z_L(t)
&\le&\frac\beta{2}\left(\frac\ell L\right)^dE
\left(\ell^{d(p-1)}\sum_{n_1,\cdots,n_{p}}
w_+^{(p)}(n_1,\cdots,n_{p};\gamma)
\med{(q_{12}^{(n_1)})^p}\right.\\\nonumber
&&\left.
-\sum_n \med{(q^{(n)}_{12})^{p}}\right).
\end{eqnarray}
Finally, one employs the property
\begin{equation}
\label{sum}
\lim_{\gamma\to0}\lim_{L\to\infty}\sum_{n_2,\cdots,n_{p}}
w_+^{(p)}(n_1,\cdots,n_{p};\gamma)= \ell^{-d(p-1)}
\end{equation}
which holds \cite{lp} since, in the Kac limit, the potential $w^{(p)}(i_1,\cdots,i_p;\gamma)$ 
becomes smoother and smoother, 
so that the sum in (\ref{sum}) converges to its Riemann integral, up to  the factor $\ell^{d(p-1)}$
which is just the size of the cell in the Riemann sum.
Therefore,
\begin{equation}
\limsup_{\gamma\to0}\limsup_{L\to\infty}\frac d{dt}\frac{-1}{\beta L^d}E \ln Z_L(t)\le0,
\end{equation}
from which (\ref{ub}) follows, after taking the limit $\ell\to\infty$.
\hfill $\Box$

{\bf Remark}
The arguments outlined above can be also employed to obtain a new proof
of the existence of the thermodynamic limit for the free energy of mean field spin glass models, independent of the
convexity argument developed in \cite{limterm} \cite{limterm2} \cite{limterm3}. 
Indeed, from the first interpolation (\ref{z1}) above it follows that
\begin{equation}
\liminf_{\gamma\to0}\lim_{L\to\infty} f_L^{(p,\gamma)}(\beta,h)\ge\limsup_{L\to\infty} f_L^{(p)}(\beta,h),
\end{equation}
while from (\ref{z2}) one obtains
\begin{equation}
\limsup_{\gamma\to0}\lim_{L\to\infty} f_L^{(p,\gamma)}(\beta,h)\le\liminf_{\ell\to\infty} f_\ell^{(p)}(\beta,h),
\end{equation}
which together imply the existence of the limit in (\ref{limtermMF}). \hfill $\Box$

\begin{lemma}
\label{lemmaLP}
For any even $p$, and any real numbers $x_1,\cdots,x_p$, one has
\begin{equation}
\frac {x_1^p+\cdots +x_p^p}p\ge x_1\cdots x_p.
\end{equation}
\end{lemma}
{\em Proof} 
One has
\begin{eqnarray}
\frac{x_1^p+\cdots +x_p^p}p\ge \left(\frac{|x_1|+\cdots +|x_p|}{p}\right)^p
\ge |x_1|\cdots |x_p|\ge x_1\cdots x_p
\end{eqnarray}
where the first inequality follows from convexity of the function
$x\to x^p$ and the second from the ``arithmetic-geometric'' inequality
\cite{gradstein}. \hfill $\Box$

\begin{lemma}
\label{sobolev}
Given numbers  $\tau_i$, $i\in \toro$, and defining 
$$
M=\frac1{L^d}\sum_{i\in\toro}
\tau_i,
$$ 
one has for $p\ge 1$
\begin{eqnarray}
\label{eqsobolev}
&&\frac1{L^d}\sum_{i_1,\cdots,i_p\in\toro}w^{(p)}(i_1,\cdots,i_p;\gamma)
\tau_{i_1}\cdots\tau_{i_p}-M^p\\\nonumber
&&=
\frac1{L^d}\sum_{r=2}^{p} M^{p-r}\left(
\begin{array}c
p\\
r\\
\end{array}
\right)
\sum_{i_1,\cdots,i_r\in\toro}w^{(r)}(i_1,\cdots,i_r;\gamma)
\prod_{s=1}^r(\tau_{i_s}-M).
\end{eqnarray}
\end{lemma}
The proof of Lemma \ref{sobolev} is postponed to Appendix A.

\section{The distribution of the local overlaps}

\label{sec:lo}

{\em Proof of Theorem \ref{teooverlap}} 

First of all, it is not difficult to generalize Theorem \ref{generalizz} to the perturbed model (\ref{hkac}),
{\em i.e.}, to prove that
\begin{eqnarray}
\label{dup}
\kac f^{(p,\gamma)}_L(\beta,\{\varepsilon\},h)=f^{(p)}(\beta,\{\varepsilon\},h)\equiv\lim_{L\to\infty}
f^{(p)}_L(\beta,\{\varepsilon\},h).
\end{eqnarray}
The additional difficulty is the presence of the $r$-spin perturbations with $r$ odd, whose effect
can be however controlled,
if the parameters $\{\varepsilon_r\}_{r\ge1}$ belong to the region $R_p$ described in Section \ref{talagrandata}. 
The proof of (\ref{dup}) is outlined in Appendix \ref{app_epsilon}.
Due to the convexity of the free energy,
the partial derivatives with respect to the perturbing parameters  exist almost 
everywhere and \cite{convex}
\begin{eqnarray}
\label{generic}
\kac \partial_{\varepsilon_{r}}f^{(p,\gamma)}_L(\beta,\{\varepsilon\},h)
=\kac\frac1{L^d}E
\med{K^{(r,\gamma)}_L}
=\lim_{L\to\infty}\partial_{\varepsilon_{r}}f^{(p)}_L(\beta,\{\varepsilon\},h),
\end{eqnarray}
where the thermal average $\med.$ corresponds to the
 Hamiltonian (\ref{hkac}) and therefore depends also on $p,\gamma$ and $\{\varepsilon\}$.
Now, using translation invariance and recalling definitions  (\ref{Hp}), (\ref{ovloc}), an immediate integration 
by parts on the Gaussian disorder shows that
\begin{eqnarray}
\frac1{L^d}E
\med{K^{(r,\gamma)}_L}=-\beta\varepsilon_{r}\left(1-E\med{(q^{\gamma}_{12})^{r}}\right).
\end{eqnarray}
On the other hand, the r.h.s. of (\ref{generic}) 
gives the same expression, only
with $q^{\gamma}_{12}$ replaced by the full overlap $q_{12}$,  and with 
the Gibbs average replaced by the Gibbs average of the mean field model (\ref{hSK}).
Therefore, one has convergence
of the moments of $P^{(p,\gamma)}_{\varepsilon}(q)$  to those of $P^{(p)}_{\varepsilon}(q)$ 
in the Kac limit, which implies (\ref{convergenza}), since the (local) overlaps are bounded random variables.
 \hfill $\Box$

 \section{From local to global order: outlook and open problems}

\label{sec:out}
The scope of this section is to discuss some perspectives opened by our work. 
The following discussion will have an informal character, with no aim to mathematical rigor. 

The main point of our work is that for small $\gamma$ the physics of  finite dimensional 
spin glasses is locally close to the one of
mean field models. The free energy tends to that of the corresponding mean field model and,
at low temperature, the Kac model exhibits locally a non-trivial distribution of the overlaps on 
length scales smaller than $\gamma^{-1}$. From the definition (\ref{ovloc}) it is clear that, given 
any two configuration $\sigma^1,\sigma^2$, the
 local overlap $q_k^\gamma$ is a smooth function of the space index $k$ on the scale 
 $\gamma^{-1}$. Indeed, one has 
 \begin{eqnarray}
 |q_k^\gamma(\sigma_1,\sigma_2)-q_l^\gamma(\sigma_1,\sigma_2)|\leq
 \sum_{i\in\toro}\frac{|\psi(\gamma|i-k|)-\psi(\gamma|i-l|)|
 }{W(\gamma)^{1/2}}\sim \gamma |k-l|. 
 \end{eqnarray}
The Kac model is therefore an example supporting the possibility, advocated several times, 
that in short range spin glasses replica symmetry breaking describes at least local ordering properties. 
The next relevant question concerns the 
possibility of long range order.  Of course local, short range, ordering does not necessary imply global, 
long range, order in which the global overlaps  
 exhibit non-trivial statistics.  In low enough dimension, for all
 positive $\gamma$, the system should be a paramagnet at all
 temperature (see  \cite{sg1d} for the discussion of the 1-dimensional case), 
where the overlap distribution is a single
 $\delta$-function in zero. This means that the typical configurations have 
overlap profiles taking locally the values in the support of the mean field overlap probability 
distribution $P^{M.F.}(q)$, 
but averaging to zero on the scale of the system size. Conversely, according to the Replica Symmetry Breaking theory (RSB) \cite{MPV}, one would expect that in high enough dimension the distribution of the local and the global overlap should coincide. This corresponds to a long range order where the overlap profiles 
dominating the overlap measure are constant in space. 

This observation suggests to define RSB in extended systems in terms of sensitiveness to ``overlap boundary 
conditions''. Take two copies of the Kac spin glass with the same quenched disorder, enclosed for simplicity
in  the 
$d$-dimensional hypercube $\Lambda_L$ of side $L$.
Consider now a region of thickness $\sim \gamma^{-1}$ around the boundary of $\Lambda_L$ and choose 
some value $p$ in the support of $P^{M.F.}(q)$. 
The boundary conditions will consist in constraining the spin configurations of the two systems so that their mutual 
local overlap $q^\gamma_k$ equals 
$p$ for all $k$ in the boundary region. 
We say that RSB long range order is present if the probability that the overlap $q_0(\sigma^1,\sigma^2)^\gamma$ around the central site 
of $\Lambda_L$ is different from $p$ vanishes in the 
thermodynamic limit. A mathematical theory of finite dimensional spin glasses should find a way to estimate this probability. 

Kac models could give the opportunity of studying this probability in a simplified setting, as it happens for the corresponding object in the ferromagnetic case. In ferromagnetic Kac systems one can define ``block 
spins'', i.e. local magnetization on scales $l$ such that $1<<l<<\gamma^{-1}$, and the probability of block spin profiles in space. This probability takes for small $\gamma$ the form of a large deviation functional \cite{cp} with rate function consisting in the space dependent mean field free-energy as a function of the profile. This is the starting point of a ``semiclassical'' 
analysis in which the saddle point treatment of the rate 
function can be used to infer the phase structure of the model for small $\gamma$, and in particular the existence of 
spontaneous magnetization and sensitivity to boundary conditions at low temperature.

In analogy with that, in the spin glass case one can introduce the probability of local overlap profiles. The formalism to study that probability within the replica method has been introduced in \cite{fpv}. In that work, an unjustified saddle point procedure was used to estimate the free-energy cost of overlap inhomogeneities, which suggested the
presence of  RSB in dimension greater or equal to $d=3$.
Straightforward application of that formalism to the Kac spin glass
model again gives a large deviation theory where the dominant profiles are constant in space. The resulting rate function is indeed similar to the one used in \cite{fpv}, thus vindicating {\it a posteriori} the saddle point procedure used in that paper. 

Of course, in order to give decisive contributions in the debate about the nature of the spin glass phase in 
finite dimensional systems, 
these arguments  should be put on a solid mathematical bases, and the large deviation approach of
\cite{fpv} should receive a justification going beyond the replica theory. Progress 
in that direction will be reported soon \cite{FDG}.

{\bf Acknowledgments}

We would like to thank T. Bodineau, M. Cassandro, F. Guerra, A. Montanari and P. Picco for illuminating discussions.
This work was supported in part by the European Community's Human Potential
programme under contract ``HPRN-CT-2002-00319 STIPCO'', and by the Swiss Science Foundation
Contract No. 20-100536/1.

\appendix

\section{Proof of Lemma \ref{sobolev}}

Eq. (\ref{eqsobolev}) follows from the following more general identity: 
for $p\ge1$ and $k=0,\cdots,p$,
\begin{eqnarray}
\nonumber
&&\frac1{L^d}\sum_{i_1,\cdots,i_p\in\toro}w^{(p)}(i_1,\cdots,i_p;\gamma)
\tau_{i_1}\cdots\tau_{i_k}(\tau_{i_{k+1}}-M)\cdots (\tau_{i_{p}}-M)
\\
\label{+gener}
=
&&M^p\delta_{k=p}+\frac1{L^d}\sum_{r=\max(2,p-k)}^{p}M^{p-r}
\left(\begin{array}c
k\\
k+r-p\\
\end{array}
\right)\\\nonumber
&&
\times
\sum_{i_1,\cdots,i_r\in\toro}w^{(r)}(i_1,\cdots,i_r;\gamma)
\prod_{s=1}^r(\tau_{i_s}-M).
\end{eqnarray}

Clearly, the identity is trivial for $k=0$, and
(\ref{eqsobolev}) follows taking $k=p$.

The proof of (\ref{+gener}) proceeds by induction on $p$ 
(for $p=1,2$ it is trivial). 
Suppose the identity is true up to $p-1$. Then, one has for $k\ge1$
\begin{eqnarray}
\label{palla}
&&
\frac1{L^d}\sum_{i_1,\cdots,i_p\in\toro}w^{(p)}(i_1,\cdots,i_p;\gamma)
\tau_{i_1}\cdots\tau_{i_k}(\tau_{i_{k+1}}-M)\cdots (\tau_{i_{p}}-M)
\\\nonumber
&&
=\frac1{L^d}\sum_{i_1,\cdots,i_p\in\toro}w^{(p)}(i_1,\cdots,i_p;\gamma)
\tau_{i_1}\cdots\tau_{i_{k-1}}(\tau_{i_{k}}-M)\cdots (\tau_{i_{p}}-M)
\\\nonumber
&&
+\frac M{L^d}\sum_{i_1,\cdots,i_{p-1}\in\toro}w^{(p-1)}
(i_1,\cdots,i_{p-1};\gamma)
\tau_{i_1}\cdots\tau_{i_{k-1}}(\tau_{i_{k}}-M)\cdots (\tau_{i_{p-1}}-M)
\\\nonumber
&&
=\frac1{L^d}\sum_{i_1,\cdots,i_p\in\toro}w^{(p)}(i_1,\cdots,i_p;\gamma)
\tau_{i_1}\cdots\tau_{i_{k-1}}(\tau_{i_{k}}-M)\cdots (\tau_{i_{p}}-M)
\\\nonumber
&&
+M^p\delta_{k=p}+
\frac M{L^d}\sum_{r=\max(2,p-k)}^{p-1}M^{p-r-1}
\left(\begin{array}c
k-1\\
k+r-p\\
\end{array}
\right)
\\\nonumber
&&
\times\sum_{i_1,\cdots,i_r\in\toro}w^{(r)}(i_1,\cdots,i_r;\gamma)
\prod_{s=1}^r(\tau_{i_s}-M),
\end{eqnarray}
where in the first step we used the property (\ref{consistenza}).
Repeating $k$ times the trick of replacing $\tau_{i_a}$
by $(\tau_{i_a}-M)+M$,
one can finally rewrite (\ref{palla}) as
\begin{eqnarray}
\nonumber
&&
M^p\delta_{k=p}+
\frac1{L^d}\sum_{i_1,\cdots,i_p\in\toro}w^{(p)}(i_1,\cdots,i_p;\gamma)
\prod_{s=1}^p(\tau_{i_s}-M)
\\\nonumber
&&
+\sum_{w=0}^{k-1} \frac 1{L^d}\sum_{r=\max(2,p-k+w)}^{p-1}M^{p-r}
\left(\begin{array}c
k-w-1\\
r-p+k-w\\
\end{array}
\right)
\\\nonumber
&&\times
\sum_{i_1,\cdots,i_r\in\toro}w^{(r)}(i_1,\cdots,i_r;\gamma)
\prod_{s=1}^r(\tau_{i_s}-M).
\end{eqnarray}
and it is not difficult to check that this coincides with the r.h.s. of 
(\ref{+gener}),
thanks to the identity
\begin{equation}
\sum_{w=0}^{k+r-p}\left(
\begin{array}c
k-w-1\\
k+r-p-w\\
\end{array}
\right)=\left(
\begin{array}c
k\\
k+r-p\\
\end{array}
\right).
\end{equation}

\hfill $\Box$

\section{Proof of Eq. (\ref{dup})}

\label{app_epsilon}

We sketch the proof of Eq. (\ref{dup}), pointing out only the differences with respect to the proof in 
Section \ref{sec:ft}.
In view of obtaining the analogue of the lower bound (\ref{lb}), 
one generalizes in an obvious way the interpolation (\ref{z1})
and obtains, in analogy with (\ref{derivata}), the identity
\begin{eqnarray}
\label{agh}
-\frac d{dt}\frac1{\beta L^d}E \ln Z_L(t)=\frac\beta2
\sum_{r=2}^p\left(\begin{array}l
p\\
r\\
\end{array}
\right)
E\Med{q_{12}^{p-r}y^r}
+\frac\beta2\varepsilon_2^2E\med{y^2}
+\frac\beta2\sum_{s\ge3}\varepsilon_s^2
\sum_{r=2}^{s} \left(\begin{array}l
s\\
r\\
\end{array}
\right)
E\Med{q_{12}^{s-r}y^r},
\end{eqnarray}
where
\begin{equation}
y=\sum_{i\in\toro}\frac{\psi(\gamma|i|)}
{W(\gamma)^{1/2}}
(\sigma^1_{i}\sigma^2_{i}-q_{12}), \;|y|\le2.
\end{equation}
All the terms in  (\ref{agh}) vanish at least as fast as $y^2$, when $y\to0$. 
This implies that, provided that
\begin{equation}
\label{condiz1}
\sum_{s\ge3}\varepsilon_s^2\sum_{r=2}^s 2^r\left(\begin{array}l
s\\
r\\
\end{array}
\right)\le \varepsilon_2^2,
\end{equation}
one has
\begin{equation}
-\frac d{dt}\frac1{\beta L^d}E \ln Z_L(t)\ge \frac\beta2
\sum_{r=2}^p\left(\begin{array}l
p\\
r\\
\end{array}
\right)
E\Med{q_{12}^{p-r}y^r}+\frac{3\beta}8 \varepsilon_2^2E\med{y^2}\ge0.
\end{equation}

As for the upper bound, the analogue of (\ref{bound_deriv_ub}) can be conveniently rewritten as
\begin{eqnarray}
\nonumber
&&-\frac d{dt}\frac1{\beta L^d}E \ln Z_L(t)=\frac{\beta}{2}\left(\frac\ell L\right)^d
E
\left(\ell^{d(p-1)}\sum_{n_1,\cdots,n_{p}}
w^{(p)}(i_{n_1},\cdots,i_{n_{p}};\gamma)
\med{q_{12}^{(n_1)}\cdots q_{12}^{(n_p)}}
-\sum_n \med{(q^{(n)}_{12})^{p}}\right)
\\\nonumber
&&+\frac\beta{2}\sum_{s\ge2}\varepsilon_s^2\left(\frac\ell L\right)^d E
\left(\ell^{d(s-1)}\sum_{n_1,\cdots,n_{s}}
w^{(s)}(i_{n_1},\cdots,i_{n_{s}};\gamma)
\med{q_{12}^{(n_1)}\cdots q_{12}^{(n_s)}}
-\sum_n \med{(q^{(n)}_{12})^{s}}\right)
+o(1),
\end{eqnarray}
where $i_n$ is the lattice site situated at the center of the $n$-th box
and the error term $o(1)$ vanishes in the Kac limit $\kac$.
Given a  set of numbers $\tau^{(n)}$, where the index $n$ runs over the cells $\Omega_n$, define the average
$\mu(\tau^{(.)})$ as
\begin{equation}
\mu(\tau^{(.)})=\frac{\sum_n \psi(\gamma |i_n|)\tau^{(n)}}{\sum_n \psi(\gamma |i_n|)}
\end{equation}
Then, it is not difficult to realize that
\begin{eqnarray}
\label{uff}
-\frac d{dt}\frac1{\beta L^d}E \ln Z_L(t)&=& -\frac\beta2 E\left\langle
\left[\mu\left((q^{(.)}_{12})^p\right)-\left(\mu\left(q^{(.)}_{12}\right)\right)^p\right]\right.\\\nonumber
&&\left.+
\sum_{s\ge2}\varepsilon_s^2\left[\mu\left((q^{(.)}_{12})^s\right)-\left(\mu\left(q^{(.)}_{12}\right)\right)^s\right]
\right\rangle+o(1).
\end{eqnarray}
Now, given a measure $\mu$ and a bounded random variable $|q|\le1$, one has for $s\ge3$,
\begin{equation}
\left|\mu(q^s)-(\mu(q))^s\right|\le \xi_s (\mu(q^2)-(\mu(q))^2),
\end{equation}
for some constant $\xi_s$ independent of $\mu$ and $q$. 
Then, it is immediate to see that, if  
\begin{equation}
\label{condiz2}
\sum_{s=3}^\infty \varepsilon_s^2\xi_s<\varepsilon_2^2,
\end{equation}
the r.h.s. of (\ref{uff}) is non-positive.
The region $R_p$ in Section \ref{talagrandata} is therefore defined by conditions (\ref{condiz1}) and 
(\ref{condiz2}).
\hfill $\Box$

\end{document}